\documentclass[twocolumn,aps]{revtex4}
\usepackage{graphicx}

\usepackage{graphicx}

\newcommand*{\be}
{\begin{equation}}
\newcommand*{\ee}
{\end{equation}}

\begin{document}
\title{Large-N transition temperature for superconducting films in a magnetic
field}
\author{L. M. Abreu and A. P. C. Malbouisson }
\address{{\it Centro Brasileiro de Pesquisas F\'{\i}sicas,
Rua Dr. X. Sigaud 150, 22290-180, Rio de Janeiro, RJ, Brazil}}
\author{J. M. C. Malbouisson and A. E. Santana}
\address{{\it Instituto de F\'{\i}sica, Universidade Federal da Bahia,
40210-340, Salvador, BA, Brazil}}

\begin{abstract}
\noindent We consider the $N$-component Ginzburg-Landau model in
the large $N$ limit, the system being embedded in an external
constant magnetic field and confined between two parallel planes a
distance $L$ apart from one another. On physical grounds, this
corresponds to a material in the form of a film in the presence of
an external magnetic field. Using techniques from dimensional and
$zeta$-function regularization, modified by the external field and
the confinement conditions, we investigate the behavior of the
system as a function of the film thickness $L$. This behavior
suggests the existence of a minimal critical thickness below which
superconductivity is suppressed.\\
\\
\noindent PACS number(s): 74.20.-z, 05.10Cc, 11.25.Hf
\end{abstract}
\maketitle

It is usually assumed that it is a good approximation to neglect
magnetic thermal fluctuations in the Ginzburg-Landau (GL) model,
when applied to study type II superconductors. This problem has
been investigated by a number of authors, both in its single
component and in its $N$- component versions. An account on the
state of the subject can be found for instance in refs.
\cite{Lawrie,Lawrie1,Brezin,Moore,Radz,Flavio}. In particular in
ref.\cite {Radz} a large-$N$ theory of a second order transition
for arbitrary dimension $D$ is presented and the fixed point
effective free energy describing the transition is found. Here we
investigate a confined version of the model studied in
ref.\cite{Radz}. We consider the vector $N$ -component
Ginzburg-Landau model in presence of an external magnetic field at
leading order in $\frac{1}{N}$, the system being submitted to the
constraint of being confined between two parallel planes a
distance $L$ apart from one another. Studies on confined field
theory have been done in the literature since a long time ago. In
particular, an analysis of the renormalization group in finite
size geometries can be found in ref.\cite {Zinn}. This study is
performed using a modified Matsubara formalism to take into
account boundary effects on scaling laws. From a physical point of
view, for $D=3$ and introducing temperature by means of the mass
term in the Hamiltonian, the model studied here should correspond
to a film-like material in presence of a magnetic field. We
investigate the behavior of the system as a function of the
separation $L$ between the planes (the film thickness), using an
extended compactification formalism in the framework of the
effective potential, introduced in recent publications \cite
{Ademir,JMario}. In \cite{adolfo} this formalism has been employed
to perform a study of the large-$N$ $\beta $-function for
superconducting films in a magnetic field. For the Ginzburg
parameter $\kappa \gg 1$ (which is the case for high temperature
superconductors) the Hamiltonian density of the GL model in an
external magnetic field can be written in the form,
\begin{equation}
{\cal H}=|(\nabla -ie{\bf A})\phi |^{2}+m_{0}^{2}|\phi
|^{2}+\frac{u}{2} |\phi |^{4},  \label{GL}
\end{equation}
where $\nabla \times {\bf A}={\bf H}$ and $m_{0}^{2}=\alpha
(T-T_{0}),$ with $\alpha >0$ and $T_{0}$ corresponding to the bulk
transition temperature in absence of external field. This
Hamiltonian density describes superconductors in the extreme type
II limit. In the following we assume that the external magnetic
field is parallel to the $z$ axis and that the gauge ${\bf
A}=(0,xH,{\bf 0})$ has been chosen. We will consider the model
(\ref{GL}) with $N$ complex components and take the large $N$
limit at $Nu$ fixed. If we consider the system in unlimited space,
the field $\phi $ should be written in terms of the well known
Landau level basis,
\begin{equation}
\phi ({\bf r})=\sum_{l=0}^{\infty }\int \frac{dp_{y}}{2\pi }\int
\frac{ d^{D-2}p}{(2\pi )^{D-2}}\hat{\phi}_{l,p_{y},{\bf p}}\chi
_{l,p_{y},{\bf p}}( {\bf r}),  \label{landau}
\end{equation}
where $\chi _{l,p_{y},p_{z}}({\bf r})$ are the Landau level eingenfunctions
given with energy eigenvalues $E_{l}(|{\bf p}|)=|{\bf p}|^{2}+(2l+1)\omega
+m_{0}^{2}$ and $\omega =eH$ is the so called cyclotron frequency. In the
above equation ${\bf p}$ is a $(D-2)$-dimensional vector.

Now, let us consider the system confined between two parallel planes, normal
to the $z$-axis, a distance $L$ apart from one another and use Cartesian
coordinates ${\bf r}=(z,{\bf z})$, where ${\bf z}$ is a $(D-3)$-dimensional
vector, with corresponding momenta ${\bf k}=(k_{z},{\bf q})$, ${\bf q}$
being a $(D-3)$ -dimensional vector in momenta space. In this case, the
model is supposed to describe a superconducting material in the form of a
film. Under these conditions the generating functional of the correlation
functions is written as,
\begin{equation}
{\cal Z}=\int {\cal D}\phi ^{\ast }{\cal D}\phi exp\left(
-\int_{0}^{L}dz\int d^{D-3}{\bf z}\;{\cal H}(|\phi |,|\nabla \phi |)\right) ,
\label{Z}
\end{equation}
with the field $\phi (z,{\bf z})$ satisfying the condition of confinement
along the $z$-axis, $\varphi (z=0,{\bf z})\;=\;\varphi (z=L,{\bf z})$. Then
the field representation (\ref{landau}) should be modified and have a mixed
series-integral Fourier expansion of the form,
\begin{eqnarray}
\phi (z,{\bf z}) &=&\sum_{l=0}^{\infty }\sum_{n=-\infty }^{\infty
}c_{n}\int \frac{dp_{y}}{2\pi }\int d^{D-3}{\bf q}\;b({\bf q})
\nonumber
\label{Fourier} \\
&&\times e^{-i\omega _{n}x\;-i{\bf q}\cdot {\bf z}}\tilde{\varphi}
_{l}(\omega _{n},{\bf q}),
\end{eqnarray}
where $\omega _{n}=2\pi n/L$, the label $l$ refers to the Landau
levels, and the coefficients $c_{n}$ and $b({\bf q})$ correspond
respectively to the Fourier series representation over $z$ and to
the Fourier integral representation over the $D-3$-dimensional
${\bf z}$-space. The above conditions of confinement of the
$z$-dependence of the field to a segment of length $L$, allow us
to proceed with respect to the $z$-coordinate, in a manner
analogous as it is done in the imaginary-time Matsubara formalism
in field theory. The Feynman rules should be modified following
the prescription,
\begin{equation}
\int \frac{dk_{z}}{2\pi }\rightarrow \frac{1}{L}\sum_{n=-\infty }^{+\infty
}\;,\;\;\;\;\;\;k_{z}\rightarrow \frac{2n\pi }{L}\equiv \omega _{n}.
\label{Matsubara}
\end{equation}
We emphasize that here we are considering an Euclidean field theory in $D$
{\it purely} spatial dimensions, we are {\it not} working in the framework
of finite temperature field theory. Temperature is introduced in the mass
term of the Hamiltonian by means of the usual Ginzburg-Landau prescription.

We consider in the following the zero external-momenta four-point function,
which is the basic object for our definition of the renormalized coupling
constant. The four-point function at leading order in $\frac{1}{N}$ is given
by the sum of all chains of single one-loop diagrams. This sum gives for the
$L$ and $\omega $ -dependent four-point function at zero external momenta at
the lowest Landau level approximation the formal expression,
\begin{equation}
\Gamma _{D}^{(4)}(0;L,\omega )=\;\frac{-u}{1+Nu\Sigma (D,L,\omega )},
\label{4-point1}
\end{equation}
where $\Sigma (D,L,\omega )$ is the Feynman integral corresponding
to the single one-loop subdiagram,
\begin{eqnarray}
\Sigma (D,L,\omega )& = & \frac{1}{L}\sum_{n=-\infty }^{\infty
}\frac{\omega }{
2\pi }  \nonumber \\
 & & \times \int \frac{d^{D-3}k}{(2\pi )^{D-3}}\;\frac{1}{\left[
k^{2}+\omega _{n}^{2}+m^{2}+\omega \right] ^{2}} .
\end{eqnarray}

The sum over $n$ and the integral over $k$ can be treated using the
formalism developed in \cite{Ademir,JMario}, which we resume below, adapted
to the situation under study. The starting point is an expression of the
form,
\begin{equation}
U=\frac{\mu ^{D-2-2s}}{(2\pi )^{2s+1}}\omega \sqrt{a}\sum_{n=-\infty
}^{+\infty }\int \frac{d^{D-3}k}{(an^{2}+c^{2}+{\bf k}^{2})^{s}},
\label{potefet1}
\end{equation}
where we have used dimensionless quantities, $c^{2}=(m^{2}+\omega )/(4\pi
^{2}\mu ^{2})\;,\;(L\mu )^{2}=a^{-1}$, and $\mu $ is a mass scale. Note that
our formalism makes sense only for dimensions $D\geq 3$. Using a well-known
dimensional regularization formula \cite{Zinn}, Eq.(\ref{potefet1}) can be
written in the form,
\begin{equation}
U=\mu ^{D-2-2s}\omega \sqrt{a}f(D,s)Z_{1}^{c^{2}}(s-\frac{D-3}{2};a),
\label{potefet2}
\end{equation}
where $f(D,s)$ is
\begin{equation}
f(D,s)=\frac{\pi ^{D-4-2s}}{2^{2s+1}}\frac{\Gamma (s-\frac{D-3}{2})}{\Gamma
(s)},  \label{fuction f}
\end{equation}
and $Z_{1}^{c^{2}}(s-\frac{D-3}{2};a)$ is one of the
Epstein-Hurwitz $zeta$ -functions defined by,
\begin{equation}
Z_{K}^{c^{2}}(u;\{a_{i}\})=\sum_{n_{1},...,n_{K}=-\infty }^{+\infty
}(a_{1}n_{1}^{2}+...+a_{K}n_{K}^{2}+c^{2})^{-u},  \label{zeta}
\end{equation}
valid for $Re(u)>\;K/2$ (in our case $Re(s)>\;(D-2)/2$). The Epstein-Hurwitz
$zeta$-function can be extended to the whole complex $s$-plane and we
obtain, after some rather long but straightforward manipulations,
\begin{eqnarray}
U &=&h(D,s)\omega \left[ \frac{1}{4}\Gamma
(\frac{2s-D+2}{2})(\frac{
m^{2}+\omega }{2\mu ^{2}})^{\frac{D-2-2s}{2}}\right.  \nonumber \\
&&\left. +\sum_{n=1}^{\infty }(\frac{\sqrt{m^{2}+\omega }}{\mu
^{2}nL})^{
\frac{D-2}{2}-s}K_{\frac{D-2}{2}-s}(nL\sqrt{m^{2}+\omega })\right]
,
\nonumber \\
&&  \label{potefet3}
\end{eqnarray}
where
\begin{equation}
h(D,s)=\frac{\mu ^{D-2-2s}}{2^{s+\frac{D-2}{2}}\pi ^{\frac{3}{2}}\Gamma (s)},
\label{h}
\end{equation}
and $K_{\nu }$ are the Bessel functions of the third kind.\newline
Applying formula (\ref{potefet3}) with $s=2$ to the integral giving $\Sigma
(D,L,\omega )$, the result is that, we can write $\Sigma (D,L,\omega )$ in
the form,
\begin{equation}
\Sigma (D,L,\omega )=\omega \left[ H(D,\omega )+G(D,L,\omega )\right] ,
\label{Sigma}
\end{equation}
where the $L$ and $\omega $ dependent contribution $G(D,L,\omega )$ comes
from the second term between brackets in Eq.(\ref{potefet3}), that is,
\begin{eqnarray}
G(D,L,\omega ) &=&\frac{1}{2^{\frac{D+2}{2}}\pi ^{\frac{3}{2}}}
\sum_{n=1}^{\infty }\left[ \frac{\sqrt{m^{2}+\omega }}{nL}\right]
^{\frac{D-6
}{2}}  \nonumber \\
&&\times K_{\frac{D-6}{2}}(nL\sqrt{m^{2}+\omega }),  \label{G}
\end{eqnarray}
and $H(D,\omega )$, is a polar parcel coming from the first term between
brackets in Eq.(\ref{potefet3}),
\begin{equation}
H(D,\omega )\propto \Gamma (2-\frac{D-2}{2})\left[ \frac{m^{2}+\omega }{2\mu
^{2}}\right] ^{\frac{D-6}{2}}.  \label{H}
\end{equation}
We see from Eq.(\ref{H}) that for even dimension $D=6$, $H(D,\omega )$ is
divergent, due to the pole of the $\Gamma $-function. Accordingly this term
must be subtracted to give the renormalized single bubble function $\Sigma
_{R}(D,L,\omega )$. We get simply,
\begin{equation}
\Sigma _{R}(D,L,\omega )=\omega G(D,L,\omega ).  \label{SigmaR}
\end{equation}
For sake of uniformity the term $H(D,\omega )$ is also subtracted in the
case of lower dimensions $D$, where no poles of $\Gamma $-functions are
present. In these cases we perform a finite renormalization. From the
properties of Bessel functions, it can be seen from Eq.(\ref{G}) that for
any dimension $D$, $G(D,L,\omega )$ satisfies the conditions,
\begin{equation}
\lim_{L\rightarrow \infty }G(D,L,\omega )=0\;,\;\;\;\;\;\;\;\
\lim_{L\rightarrow 0}G(D,L,\omega )\rightarrow \infty ,  \label{GG1}
\end{equation}
and $G(D,L)>0$ for any values of $D$, $L$.

Taking inspiration from Eq.(\ref{4-point1}), let us define the $L$
and $ \omega $ dependent {\it renormalized} coupling constant
$u_{R}(D,L,\omega )$ at the leading order in $1/N$ as,
\begin{eqnarray}
\Gamma _{D,R}^{(4)}(0,L,\omega ) &\equiv &-u_{R}(D,L,\omega )=  \nonumber \\
&=&\frac{-u}{1+Nu\Sigma _{R}(D,L,\omega )}.  \label{lambR}
\end{eqnarray}
From Eqs.(\ref{lambR}) and (\ref{SigmaR}) we can write the new $L$
and $ \omega $ dependent renormalized coupling constant,
\begin{eqnarray}
\omega N u_{R}(D,L,\omega ) &\equiv &\omega \beta _{R}(D,L,\omega
)\equiv
g(D,L,\omega )  \nonumber \\
&=&\frac{\omega \beta }{1+\omega \beta G(D,L,\omega )}.
\label{lambdaR1}
\end{eqnarray}
We see from Eq.(\ref{GG1}) that $\beta =Nu$ corresponds to the
renormalized coupling constant in absence of boundaries and of
external field. Note that since the coupling constant $\beta $ has
mass dimension of $4-D$, then the coupling constant $g(D,L,\omega
)$\ has mass dimension of $6-D$.

In order to study the critical behavior of our system, we start
from the gap equation in absence of external field and of
boundaries, in the disordered phase,
\begin{equation}
\xi ^{-2}=m_{0}^{2}+\frac{V(N+2)}{N}\int \frac{d^{D}k}{ (2\pi
)^{D}}\;\frac{1}{{k^{2}+\xi ^{-2}}},  \label{gap0}
\end{equation}
where $\xi $ is the correlation length. In our case, in particular
in the neighborhood of the critical curve, the gap equation
reduces to a ($L$,$\omega$)-dependent Dyson-Schwinger equation.
So, the generalization to our case of Eq.(\ref {gap0}) in the
neighborhood of criticality can be written in the form,
\begin{eqnarray}
\xi ^{-2} &=&m_{0}^{2}+\omega +\frac{g(D,L,\omega
)}{2^{\frac{D-2}{2}}\sqrt{
\pi }}\frac{(N+2)}{N}  \nonumber \\
&&\times \frac{1}{L}\sum_{n=-\infty}^{\infty }\int \frac{d^{D-3}k}{\left( 2\pi
\right) ^{D-2}}\frac{1}{{\bf k}{^{2}+}\omega _{n}^{2}{+\xi ^{-2}}}.
\label{massaR11}
\end{eqnarray}
In Eq.(\ref{massaR11}), we remember that $\xi ^{-2}=m^{2}(L,\omega
)+\omega $ (the pole of the propagator of $\phi $ in presence of a
magnetic field is at $m^{2}(L,\omega )=-\omega
$)\cite{Lawrie,Lawrie1} and $g(D,L,\omega )$ is the renormalized $
(L,\omega$)-dependent coupling constant, which is itself a
function of $\xi ^{-2}$ via the mass $m(L,\omega )$. Performing
steps analogous to those leading from Eq.(\ref{potefet1}) to
Eq.(\ref{potefet3}), Eq.(\ref {massaR11}) becomes
\begin{eqnarray}
\xi ^{-2} &=&m_{0}^{2}+\omega +\frac{g(D,L,\omega
)}{2^{\frac{D}{2}}\pi ^{
\frac{3}{2}}}\frac{(N+2)}{N}  \nonumber \\
&&\times \sum_{n=1}^{\infty }\left[ \frac{\xi ^{-1}}{nL}\right]
^{\frac{D-4}{ 2}}K_{\frac{D-4}{2}}(nL\xi ^{-1}).  \label{massaR1}
\end{eqnarray}

The coupling constant $g(D,L,\omega )$ is given by Eq.(\ref{lambdaR1}) with
$G(D,L,\omega )$ in Eq.(\ref
{G}) replaced by,
\begin{equation}
G(D,L,\omega )=\frac{1}{2^{\frac{D}{2}+1}\pi ^{\frac{3}{2}}}
\sum_{n=1}^{\infty }\left[ \frac{\xi ^{-1}}{nL}\right]
^{\frac{D-6}{2}}K_{ \frac{D-6}{2}}(nL\xi ^{-1}).  \label{G1}
\end{equation}
Eqs.(\ref{massaR1}), (\ref{lambdaR1}) and (\ref{G1}) are in fact a
complicated set of integral equations involving $\xi ^{-2}$, since
$g(D,L,\omega )$ is also dependent on the $L$ and $\omega $
inverse correlation length.

If we limit ourselves to the neighborhood of criticality, $\xi
^{-2}\approx 0$, we may investigate the behavior of the system by
using in both Eqs.(\ref {lambdaR1}) and (\ref{G1}) and in
Eq.(\ref{massaR1}) an asymptotic formula for small values of the
argument of Bessel functions,
\begin{equation}
K_{\nu }(z)\approx \frac{1}{2}\Gamma (\nu )\left( \frac{z}{2}\right) ^{-\nu
},\;\;\;(z\sim 0),  \label{K}
\end{equation}
which allows after some straightforward manipulations, in the
large $N$ -limit, to write Eq.(\ref{massaR1}) in the form
\begin{equation}
m_{0}^{2}+\omega _{c}+\frac{g(D,L,\omega _{c})}{8\pi
^{\frac{3}{2}}} \Gamma (\frac{D}{2}-2)L^{4-D}\zeta (D-4)\approx 0,
\label{mDysoncr}
\end{equation}
where $\zeta (D-4)$ is the Riemann $zeta$-function, $\zeta
(D-4)=\sum_{n=1}^{\infty }(1/n^{D-4})$, defined for $D>5$ and we have used the
label $c$ to indicate that we are in the region of criticality). Similarly
inserting Eq.(\ref{K}) in Eqs.(\ref{lambdaR1}) and (\ref{G1}) $g(D,L,\omega )
$ can be written for $\xi ^{-2}\approx 0$ as,
\begin{equation}
g(D,L,\omega _{c})\approx \frac{\omega _{c}\beta }{1+\beta \omega
_{c}A(D,\mu )L^{6-D}\zeta (D-6)},  \label{lambdaR2}
\end{equation}
where $A(D,\mu )=\frac{1}{32\pi ^{\frac{3}{2}}}\Gamma (\frac{D-6}{2})$.

We can not obtain critical lines in dimension $D\leq 5$ by a
limiting procedure from Eq.(\ref{mDysoncr}). For $D=3$, which
corresponds to the physically interesting situation of the system
confined between two parallel planes embedded in a $3$-dimensional
Euclidean space, we can obtain critical lines, by performing an
analytic continuation of $\zeta (z)$ to values of the argument
$z\leq 1$, by means of the reflection property of $zeta
$-functions,
\begin{equation}
\zeta (z)=\frac{1}{\Gamma (z/2)}\Gamma (\frac{1-z}{2})\pi
^{z-\frac{1}{2} }\zeta (1-z),  \label{extensao}
\end{equation}
which defines a meromorphic function having only one simple pole
at $z=1$. We obtain taking $m_{0}^{2}=\alpha (T-T_{0})$, for $D=3$
(the physical dimension), the critical surface
\begin{equation}
\alpha (T_{c}-T_{0})+\omega _{c}+\frac{1}{8\pi ^{3}}g(D=3,L,\omega
_{c})L\zeta (2)=0.  \label{critica}
\end{equation}
Using Eqs.(\ref{lambdaR2}) and (\ref{extensao}) to evaluate $g(D=3,L,\omega
_{c})$, the tabulated values to the several $\Gamma $ and $\zeta $ functions
appearing in the above formulas, we obtain
\begin{equation}
\alpha (T_{c}-T_{0})+\omega _{c}+\frac{60\beta \omega
_{c}L}{2880\pi +\beta \omega _{c}L^{3}}=0,  \label{critica1}
\end{equation}
$\alpha$ and $\beta$ being the phenomenological Ginzburg-Landau
parameters. Notice that as $L\rightarrow \infty$ we obtain from
Eq. (\ref{critica1}) the critical line $\alpha
(T_{c}-T_{0})+\omega _{c}=0$ for the bulk.

In terms of the dimensionless quantities, respectively reduced
critical field, temperature and film thickness, $h=\omega_c
\xi^2_0$, $t=T_c/T_0$ and $l=L/\xi_0$, where $\xi_0=(\alpha
T_o)^{-1/2}$ is the zero-temperature Ginzburg-Landau coherence
length, the above equation can be rewritten as
\begin{eqnarray}
h(l,t) &=&\frac{1}{2Bl^{3}}\left\{ -60Bl-Btl^{3}+Bl^{3}-2880\pi \right. \nonumber \\
&&+\left. \left[ (60Bl+Btl^{3}-Bl^{3}+2880\pi )^{2}\right. \right. \nonumber \\
&&+\left. \left. 11520\pi Bl^{3}(1-t)\right] ^{1/2}\right\} ,
\label{htl}
\end{eqnarray}
where $B=\beta \xi_0$. The surface $h=h(l,t)$ is illustrated in
Fig. 1. We recall that, since we have used the lowest Landau level
approximation in our calculations, this surface is only meaningful
for high values of the external field, that is for low
temperatures and thick films.

\begin{figure}[h]
\includegraphics[{height=8.0cm,width=8.0cm}]{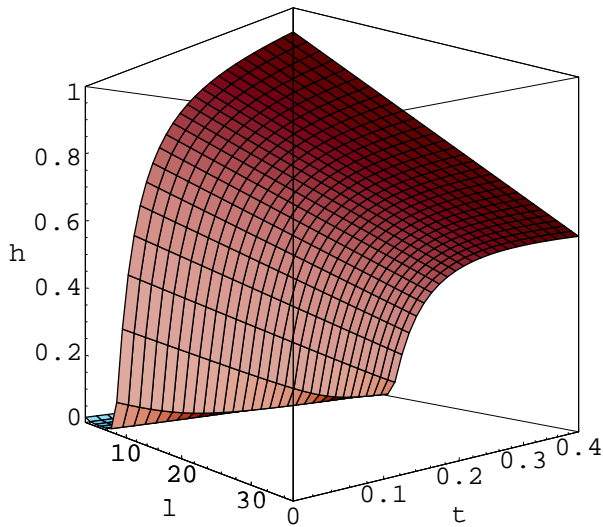}
\caption{Plot of the surface $h=h(l,t)$ as defined by Eq.
(\ref{htl}), taking $B=10^4$.}
\end{figure}

We can see from Fig. 1 that each value of $l$ defines a critical
line on the $h \times t$ plane, corresponding to a film of
thickness $L$. This set of critical lines suggests the existence
of a minimal value for the thickness $L$ below which
superconductivity is suppressed. Indeed, this can be seen from the
plot of the reduced critical field at zero temperature, $h_0$, as
a function of the inverse of the reduced film thickness, shown in
Fig. 2. This behavior may be contrasted with the linear decreasing
of $T_{c}$ with the inverse of the film thickness in absence of
external field that has been found experimentally in materials
containing transition metals, for example, in Nb \cite{Nb3} and in
W-Re alloys \cite{W-Re}; for these cases, it has been explained in
terms of proximity, localization and Coulomb-interaction effects.
Notice, however, that our results do not depend on microscopic
details of the material involved nor account for the influence of
manufacturing aspects, like the kind of substrate on which the
film is deposited. In other words, our results emerge solely as a
topological effect of the compactification of the Ginzburg-Landau
model in one direction.

\begin{figure}[t]
\includegraphics[{height=5.0cm,width=8.0cm}]{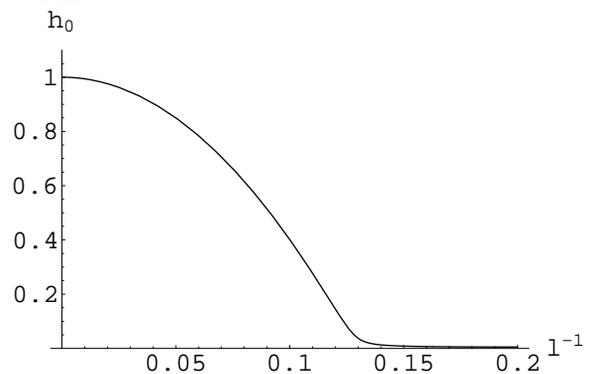}
\caption{Zero-temperature reduced critical field as a function of
the inverse of the reduced film thickness, obtained from Eq.
(\ref{htl}) taking $t=0$ and $B=10^4$.}
\end{figure}

This work was partially supported by CNPq, Brazil.


\begin{references}
\bibitem{Lawrie} I. D. Lawrie, Phys. Rev. B {\bf 50}, 9456 (1994).
\bibitem{Lawrie1}  I. D. Lawrie, Phys. Rev. Lett. {\bf 79}, 131 (1997).

\bibitem{Brezin}  E. Br\'{e}zin, D. R. Nelson, and A. Thiaville, Phys. Rev.
B {\bf 31}, 7124 (1985).

\bibitem{Moore}  M. A. Moore, T.J. Newman, A. J. Bray, S. -K. Chin, Phys. Rev. B
{\bf 58}, 936 (1998).

\bibitem{Radz}  L. Radzihovsky, Phys. Rev. Lett. {\bf 74}, 4722 (1995); {\it
ibid.} {\bf 76}, 4451 (1996); I. F. Herbut and Z. Tesanovic, {\it ibid.}
{\bf 76}, 4450 (1996).

\bibitem{Flavio}  C. de Calan, A. P. C. Malbouisson, and F. S. Nogueira,
Phys. Rev. B {\bf 64}, 212502 (2001).

\bibitem{Zinn}  J. Zinn-Justin, {\it Quantum Field Theory and Critical
Phenomena} (Clarendon Press, Oxford, 1996).

\bibitem{Ademir}  A. P. C. Malbouisson, J. M. C. Malbouisson, A. E. Santana,
Nucl. Phys. B {\bf 631}, 83 (2002).

\bibitem{JMario}  A. P. C. Malbouisson, J. M. C. Malbouisson, J. Phys. A: Math.
Gen. {\bf 35}, 2263 (2002).

\bibitem{adolfo}  A. P. C. Malbouisson, Phys. Rev. B {\bf 66}, 092502
(2002).

\bibitem{Elizalde}  A. Elizalde, E. Romeo, J. Math. Phys. {\bf 30}, 1133
(1989).

\bibitem{Nb3}  M. S. M. Minhaj, S. Meepagala, J. T. Chen, and L. E. Wenger,
Phys. Rev. B {\bf 49}, 15235 (1994).

\bibitem{W-Re}  H. Raffy, R. B. Laibowitz, P. Chaudhari, and S. Maekawa,
Phys. Rev. B {\bf 28}, R6607 (1983).
\end{references}
\end{document}